\documentclass[aps, prd,  onecolumn, preprintnumbers, nofootinbib,11pt]{revtex4-1}
\pdfoutput=1
\usepackage{myoptions}

\begin{document}

\title{Entanglement entropy across a deformed sphere}

\author{M\'ark Mezei}
\affiliation{
Princeton Center for Theoretical Science,\\ Princeton University, Princeton, NJ 08544}


\begin{abstract}

\noindent I study the entanglement entropy (EE) across a deformed sphere in conformal field theories (CFTs). I show that the sphere (locally) minimizes the universal term in EE among all shapes. In~\cite{Allais:2014ata} it was derived that the sphere is a local extremum, by showing that the contribution linear in the deformation parameter is absent. In this paper I demonstrate that the quadratic contribution is positive and is controlled by the coefficient of the stress tensor two-point function, $C_T$. Such a minimization result contextualizes the fruitful relation between the EE of a sphere and the number of degrees of freedom in field theory. I work with CFTs with gravitational duals, where all higher curvature couplings are turned on. These couplings parametrize conformal structures in stress tensor $n$-point functions, hence I show the result for infinitely many CFT examples.

\end{abstract}

\maketitle

\section{Introduction and summary of results}

\subsection{Introduction}

\begin{figure}[!h]
\begin{center}
\includegraphics[scale=0.4]{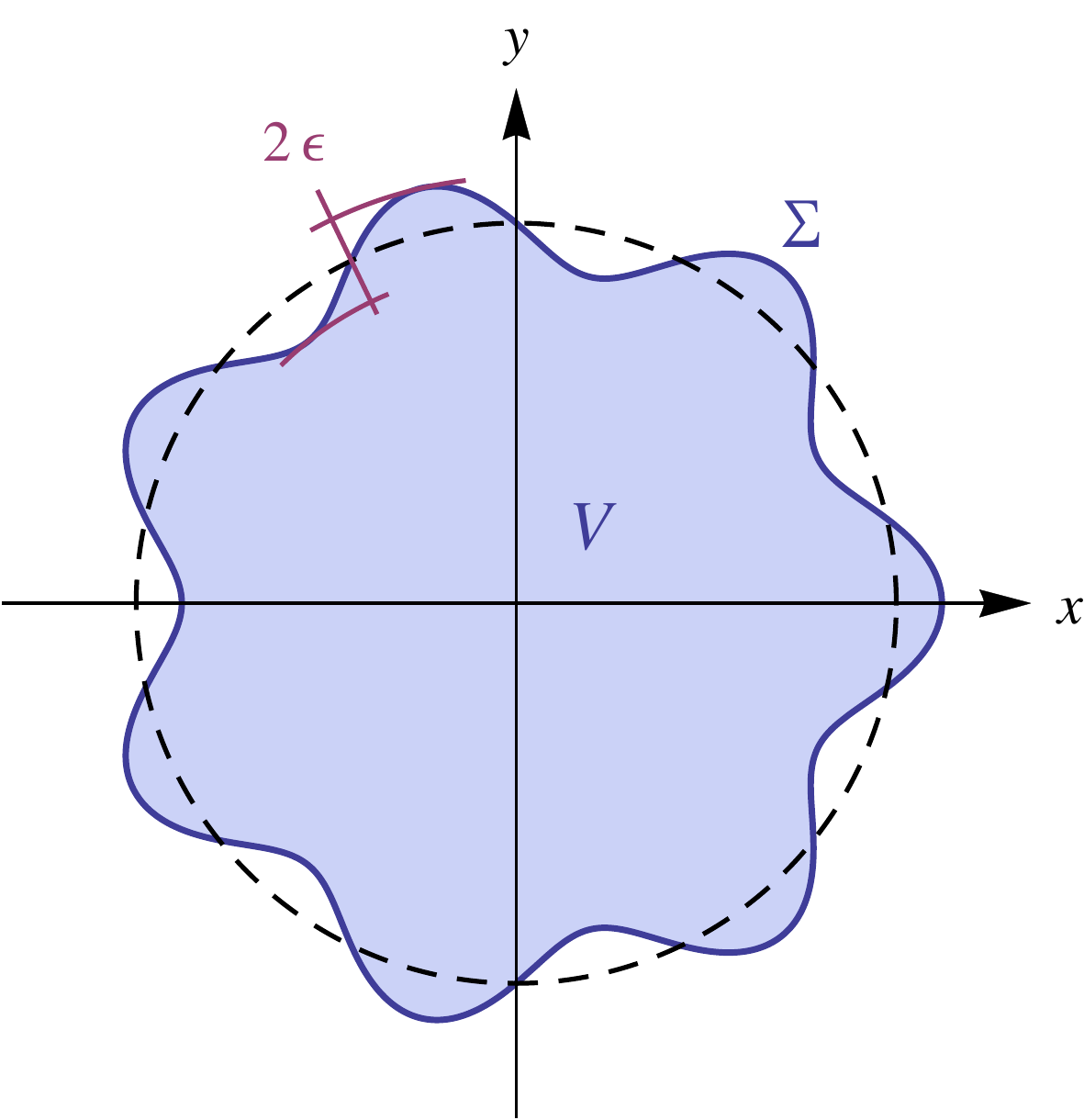}
\end{center}
\caption{\label{fig:wiggly_region} A $d=3$ example of an entangling surface $\Sig$, which is a deformed circle~\eqref{GeomRegion}. EE is the entropy of the reduced density matrix of region $V$, $\Sigma = \partial V$. The change of universal term in EE, $s_\text{univ}$ due to a deformation of amplitude $\ep$ is $\sO(\epsilon^2$) and is given by  \eqref{ShapeDep} in a CFT$_3$.}
\end{figure}

Entanglement entropy (EE), a quantity associated to subregions in quantum systems, plays an increasingly important role in recent developments in a wide range of fields; see~\cite{Ryu:2006ef,Amico:2007ag,Calabrese:2009qy,Casini:2009sr} for some reviews from this broad spectrum. In this paper, I investigate how the ground state EE of a conformal field theory (CFT) depends on the shape of the entangling surface $\Sig$. See Fig.~\ref{fig:wiggly_region} for an example of an entangling surface of interest to this work.

In~\cite{Allais:2014ata} it was conjectured that in the ground state of a CFT living in flat space, the sphere minimizes the universal contribution to EE in all dimensions among all shapes of sphere topology.\footnote{I thank Eric Perlmutter for discussions about the topology of $\Sig$.} In this paper I show that the universal term for a deformed sphere has the shape dependence
\es{ShapeDep}{
s(\Sig)_\text{univ}=s_d^{\text{(sphere)}}+\ep^2 \, C_T\,F[\text{shape}]+\sO(\ep^3) \,,
}
where $s_d^{\text{(sphere)}}$ is the contribution of the sphere, $\ep$ is the amplitude of the deformation, $C_T$ is the overall coefficient of the stress tensor two-point function~\eqref{TT2Point}, and $F[\text{shape}]$ is a positive semi-definite functional of the shape of the surface that I determine explicitly. The term linear in $\ep$ was shown to be absent in~\cite{Allais:2014ata}.
Because $C_T$ is positive in a unitary theory, I have shown that the sphere is a local minimum among all shapes for the universal piece in EE.\footnote{Proving that the sphere is a global minimum would likely require new methods.} 

There is a fruitful relation between the EE of a sphere and the number of degrees of freedom in field theory. The Wilsonian renormalization group flow between two CFT fixed points leads to the loss of degrees of freedom. This intuition is captured by the inequality
\es{CTheorem}{
\le[s_d^{\text{(sphere)}}\ri]_\text{UV}\geq \le[s_d^{\text{(sphere)}}\ri]_\text{IR} \,,
}
where the subscripts UV and IR denote  $s_d^{\text{(sphere)}}$ at the ultraviolet and infrared CFT. This inequality (better known as C-theorem) is know to be true in $d=2,3,4$~\cite{Zamolodchikov:1986gt,Cardy:1988cwa,Casini:2004bw,Myers:2010tj,Komargodski:2011vj,Casini:2012ei}. 

From the EE perspective it is not a priori clear why the number of degrees of freedom corresponds to $\Sig=S^{d-2}$ and not some other shape.\footnote{In a CFT $s_d^{\text{(sphere)}}$ is equivalent to the universal part of the Euclidean sphere free energy~\cite{Casini:2011kv}. There does not exist such a connection for other shapes. This property picks out $s_d^{\text{(sphere)}}$ as a special quantity.}  The sphere is picked out from among all nearby shapes by the appealing minimization result~\eqref{ShapeDep}, and also by the fact that it is the only shape whose EE counts the degrees of freedom; as argued below. In $d=4$ it was shown in~\cite{Anselmi:1997ys} that no linear combination of $s_d^{\text{(sphere)}}$ and $C_T$\footnote{In more familiar terms, $a_4$ and $c_4$.} obeys monotonicity~\eqref{CTheorem}. In $d=3,5$ it is also known that $C_T$ does not obey a C-theorem~\cite{Nishioka:2013gza,Fei:2014yja}. Combining these results with~\eqref{ShapeDep} makes it unlikely (in $d=4$ impossible) that any deformed sphere obeys a C-theorem.

I use holography to obtain my results. The holographic examples I study are arbitrary higher curvature gravity theories, whose Lagrangians do not contain derivatives of the curvature tensor. Through AdS/CFT I expect that these theories capture a large range of possible behaviors for  stress tensor $n$-point functions in CFTs. From field theory arguments it is clear that these are the correlators that control the shape dependence of EE~\cite{Rosenhaus:2014woa}. I only use higher curvature gravity in the above kinematical sense. I also work only to linear order in the higher curvature couplings.\footnote{See Sec.~\ref{sec:FieldTheory} for further discussions of my use of the holographic setup. See e.g.~\cite{Hung:2011nu,Safdi:2012sn,Hung:2014npa}  for works that use higher derivative gravity with a similar philosophy.} In these class of theories I get the same answer~\eqref{ShapeDep}. Hence, my results are valid in infinitely many dual CFT examples.

\subsection{Summary of results}

I continue the discussion with some more details. EE is divergent in the continuum limit. The leading divergence is the well-known area law~\cite{Bombelli:1986rw,Srednicki:1993im}. The divergence structure of EE in a $d$ dimensional CFT is (see e.g.~\cite{Liu:2012eea}):
\es{Universal}{
S(\Sig)_{EE}&={R^{d-2}\ov \de^{d-2}}+{R^{d-4}\ov \de^{d-4}}+\dots +\begin{cases}
{R\ov \de}+(-1)^{d-1\ov 2} s(\Sig)_\text{univ}+\dots \qquad &\text{$d$ odd,}\\
{R^2\ov \de^2}+(-1)^{d-1\ov 2} s(\Sig)_\text{univ}\log {R\ov\de}+\dots \qquad &\text{$d$ even,}
\end{cases}
}
where $R$ is the characteristic size of the entangling region $\Sig$. To avoid clutter I omitted the coefficients of the non-universal pieces. I am interested in the shape dependence of the universal piece $s(\Sig)_\text{univ}$. It is universal in the sense that two calculations of EE done in different regularization schemes will give the same result for $s(\Sig)_\text{univ}$, while the coefficients of the other terms in~\eqref{Universal} will not agree. 

In this paper I restrict my attention to the case when $\Sig$ is a deformed sphere (see Fig.~\ref{fig:wiggly_region} for a $d=3$ example). I describe $\Sig$ in polar coordinates as
\es{GeomRegion}{
r(\Omega_{d-2})=1+\ep \sum_{\ell, m_1,\dots, m_{d-3}} a_{\ell, m_1,\dots, m_{d-3}} Y_{\ell, m_1,\dots, m_{d-3}}(\Omega_{d-2}) \,,
} 
where $\ep$ is a small parameter, $\Omega_{d-2}$ are the coordinates on $S^{d-2}$, and I decomposed the deformation into (real) hyper-spherical harmonics that are eigenfunctions of the Laplacian on $S^{d-1}$.\footnote{They satisfy the equation:
\es{LaplacianEigen}{
\Delta Y_{\ell, m_1,\dots, m_{d-3}}(\Omega_{d-1}) =-\ell(\ell+d-3) Y_{\ell, m_1,\dots, m_{d-3}}(\Omega_{d-1}) \,.
}
I choose a normalization, in which the norm of $Y$ is one. This means that in $d=3$:
\es{Y3d}{
Y_\ell^{(c)}(\Phi)={1\ov\sqrt{\pi}}\cos(\ell\Phi)\,, \qquad
Y_\ell^{(s)}(\Phi)={1\ov\sqrt{\pi}}\sin(\ell\Phi)\,.
}}

For such a deformed sphere EE has a small $\ep$ expansion. I am interested in the universal piece:
\es{sExpansion}{
s(\Sig)_\text{univ}&=s_d^{\text{(sphere)}}+\ep^2 s_d^{(2)}(\Sig)+\sO(\ep^3) \,,
}
where the piece linear in $\ep$ vanishes~\cite{Allais:2014ata}, and $s_d^{(2)}(\Sig)$ will be my main focus. In this paper, for an infinite family of holographic examples I demonstrate that
\es{EinsteinArea}{
s_d^{(2)}(\Sig)&=C_T\,  {\pi^{d+2\ov2}(d-1)\ov 2^{d-2} \Gamma(d+2)\Gamma\le(d\ov2\ri)}\sum_{\ell, m_1,\dots, m_{d-3}} a_{\ell, m_1,\dots, m_{d-3}}^2 \prod_{k=1,\dots d} (\ell+k-2)\times \begin{cases}
{\pi\ov 2} \qquad &\text{$d$ odd,}\\
1 \qquad &\text{$d$ even,}
\end{cases}
}
where $C_T$ is determined by the stress tensor two-point function~\eqref{TT2Point}.\footnote{In~\cite{Nozaki:2013vta} an equivalent integral expression is given for the shape dependence of EE in the case of Einstein gravity. In $d=3$ the formula is
\es{3dResult}{
s_3^{(2)}(\Sig)&={\pi^3 \,C_T\ov 24}\sum_{\ell}\le((a_\ell^{(c)})^2+(a_\ell^{(s)})^2\ri)(\ell^3-\ell)\,,
}
in agreement with the Einstein gravity result of~\cite{Allais:2014ata}, if I take into account the normalization of $Y_\ell^{(c,s)}$ defined in~\eqref{Y3d}.} This is the main technical result of the paper.   Although I think I gather overwhelming evidence for this formula in this paper, it would be nice to prove~\eqref{EinsteinArea} from CFT calculations.\footnote{One caveat is that in $d=3$ in theories that violate parity, there is an interesting new structure in the stress tensor three-point function that might contribute to $s_3^{(2)}(\Sig)$, while all my examples are parity preserving. See Sec.~\ref{sec:FieldTheory} for more discussion.}

\pagebreak

An important aspect of~\eqref{EinsteinArea} is that it comes from a local term in even $d$, but a nonlocal term in odd $d$. This is seen from the highest power of $\ell$ appearing in~\eqref{EinsteinArea}, which is $\ell^d$. In even dimensions the universal term in EE multiplies a logarithmic divergence, and hence its shape dependence is given by a local functional of the geometric invariants of $\Sigma$, as I discuss in more detail in Sec.~\ref{sec:FieldTheory}. In particular it is a even functional of the extrinsic curvature. Accordingly, in~\eqref{EinsteinArea} the highest power of $\ell$ is even.
In general, if $s_\text{univ}$ were to be a local functional, the highest power of $\ell$ in its expansion would be even, as a local $s_\text{univ}$ has to be an even functional of the extrinsic curvature of $\Sig$~\cite{Grover:2011fa,Liu:2012eea}.\footnote{This is a consequence of the definition of EE: in a pure state EE is the same for a region $V$ and its complement $\bar{V}$. Because the extrinsic curvature changes sign, $K\to-K$ under the change of orientation of $\Sig$, the functional has to be even in $K$.} In odd dimensions, the highest power is odd, thus $s_\text{univ}$ has to be a non-local functional of the geometric quantities of $\Sig$.

Note also that $\ell=0,1$ deformations give zero contribution to~\eqref{EinsteinArea}. This had to be the case, as $\ell=0$ is a dilation, while $\ell=1$ (to leading order) is a translation of the sphere, neither of which should change the value of $s_\text{univ}$.

The rest of this paper presents a derivation of~\eqref{EinsteinArea}, organized as follows. In Sec.~\ref{sec:FieldTheory} I discuss how this work is related to recent progress in the perturbative approach to EE, and show that~\eqref{EinsteinArea} agrees with formulae available in the literature for $d=4,6$. In Sec.~\ref{sec:Gravity} I first review results of~\cite{Dong:2013qoa,Sen:2014nfa}, then using these results I derive~\eqref{EinsteinArea} for an infinite family of higher curvature gravity theories.

\section{Relation to field theory results}\label{sec:FieldTheory}

\subsection{Field theory results for EE}

In $d=4,6$~\eqref{EinsteinArea} reproduces the appropriate expansions of known results~\cite{Ryu:2006ef,Solodukhin:2008dh,Safdi:2012sn}. In $d=4$ Solodukhin's formula~\eqref{genes} determines the universal piece for all CFTs~\cite{Ryu:2006ef,Solodukhin:2008dh}:
\es{genes}{
s_\text{univ} &= { a_4\ov 180} \int_\Sig d^2 \sig \sqrt{\ga} \, E_2 + {c_4\ov 240\pi} \int_\Sig d^2 \sig \sqrt{\ga} \, I_2\,,\\
 I_2 &=    K_{ab} K^{ab} - \ha K^2  \,,
}
 where $a_4$ and $c_4$ are coefficients of the trace anomaly,\footnote{I normalize $a_4$ and $c_4$ so that they both equal one for a real scalar field.} $E_2$ is the Euler density normalized such that $\int_{S^2} d^2 \sig \sqrt{\ga} \, E_2=2$,  $\ga$ is the induced metric, and $K$ the extrinsic curvature.
 
By plugging~\eqref{GeomRegion} into~\eqref{genes} after a short calculation I obtain:
\es{SolodukhinExp}{
s_4^{(2)}(\Sig)&={c_4\ov 480\pi} \sum_{\ell, m} a_{\ell, m}^2 \prod_{k=1,\dots 4} (\ell+k-2) \,.
}
Using that in my normalization $C_T={1\ov  3\, \pi^4}\, c_4$, I get a precise match with~\eqref{EinsteinArea}. 
 
Because the first term in~\eqref{genes} is topological, shapes continuously connected to $S^2$ give the same contribution. It is easy to see that $I_2$ is nonnegative and vanishes only for the sphere. This shows that the sphere minimizes the universal term in EE among all topologically equivalent shapes~\cite{Astaneh:2014uba}.

In $d=6$ the analog of~\eqref{genes} is known up to some undetermined numerical coefficients~\cite{Safdi:2012sn}. To $\sO(\ep^2)$ only one term contributes, whose coefficient is fixed. In the notation of~\cite{Safdi:2012sn}:
\es{SafdiExp}{
s_d^{(2)}(\Sig)&=6\pi B_3 T_3\big\vert_{\sO(\ep^2)}=6\pi B_3  \sum_{\ell, m} a_{\ell, m}^2 \prod_{k=1,\dots 6} (\ell+k-2) \,,
}
where $T_3\big\vert_{\sO(\ep^2)}$ is the $\sO(\ep^2)$ piece of $T_3$ and $B_3$ is one of the coefficients of the trace anomaly. From~\cite{Bastianelli:2000hi,Lewkowycz:2014jia} it is known that $B_3={\pi^3\ov 193536}\, C_T$. Then~\eqref{SafdiExp} agrees with~\eqref{EinsteinArea}. These field theory checks increase my confidence in~\eqref{EinsteinArea}.

\subsection{Relation to the perturbative approach to EE}

Rosenhaus and Smolkin recently initiated a perturbative approach to the reduced density matrix~\cite{Rosenhaus:2014woa}, which immediately translates to a perturbative approach to R\'enyi entropy~\cite{Lewkowycz:2014jia}.\footnote{See however~\cite{Lewkowycz:2013laa,Lee:2014zaa} for subtleties arising for the simplest CFT, the free scalar.} One can get EE from R\'enyi entropies $S_q$ by analytic continuation:
\es{Renyi}{
S_{EE}&=\lim_{q\to 1} S_q \,.
}
To first order in perturbation theory, this continuation is straightforward, and the result is just the first law of entanglement~\cite{Bhattacharya:2012mi,Blanco:2013joa,Wong:2013gua}. The perturbation theory for R\'enyi entropies remains sound at all orders, although technical difficulties may arise due to the need to evaluate integrated $n$-point functions. However, one may worry that the analytic continuation of these results to EE may be subtle.\footnote{I thank Xi Dong for discussions on this point.} For shape deformations at $\sO(\ep^2)$ (or equivalently, deformations of the background metric~\cite{Banerjee:2011mg,Rosenhaus:2014woa,Allais:2014ata}) the continuation results in an integrated stress tensor three-point function~\cite{Rosenhaus:2014woa,Lewkowycz:2014jia}.
\cite{Rosenhaus:2014zza} recently tried to evaluate this integral in $d=4$ and failed to reproduce the known result,~\eqref{genes} appropriately expanded. With these confusing state of affairs, the result~\eqref{EinsteinArea} may prove to be a useful light post.

From the field theory results described above, it is clear that stress tensor correlators control the shape dependence of EE. I will now describe some facts about stress tensor two- and three-point functions.
The stress tensor two-point function takes the form:
\es{TT2Point}{
\<\, T_{\mu\nu}(x)\, T_{\rho\lambda}(0)\>&={C_T \ov x^{2d}}\,\le[\frac12 \le(I_{\mu\rho} I_{\nu\lambda}+I_{\mu\lam} I_{\nu\rho}\ri)-{\delta_{\mu\nu} \delta_{\rho\lambda} \ov d}\ri]\,, \\
I_{\mu\nu}&\equiv \delta_{\mu\nu} -2 {x_\mu x_\nu\ov x^2} \,.
}
The three-point function is determined by three coefficients~\cite{Osborn:1993cr,Erdmenger:1996yc}.\footnote{In $d=2$ there is only one, in a $d=3$ parity invariant theory only two independent coefficients. In a parity violating theory in $d=3$ a new structure was found in~\cite{Maldacena:2011nz,Giombi:2011rz}.} Instead of writing down the three-point function, I choose an integrated version of it, the energy one-point function in a state created by the insertion of the stress tensor with traceless polarization tensor, which only has indices in the spatial directions,  $\ep_{ij}$~\cite{Hofman:2008ar,Buchel:2009sk}. It is given by 
\es{Hofman}{
\<{\cal E}(\vec{n})\>=\frac{E}{\vol(S^{d-2})}\le[1+t_2\le(\frac{\e_{ij}^* \e_{ik}n^jn^k}{\e_{ij}^* \e_{ij}}-\frac{1}{d-1}\ri)+t_4 \le(\frac{|\e_{ij}n^jn^k|^2}{\e_{ij}^*\e_{ij}}-\frac{2}{d^2-1}\ri)\ri]\,,
}
where $\vec{n}$ picks the direction on $S^{d-2}$, $E$ is the energy of the state, and $C_T,\, t_2,\, t_4$ are the constants that determine the stress tensor three-point functions.
\footnote{Note that in $d=3$ the term multiplying $t_2$ is zero, correspondingly there are just two independent coefficients in the stress tensor three-point function in a parity preserving theory.
} 
\footnote{The relation between $C_T,\, t_2,\, t_4$ and the CFT coefficients ${\mathcal{A}}, \ {\mathcal{B}}, \  {\mathcal{C}}$ of~\cite{Erdmenger:1996yc} are 
\begin{align}
\begin{split}
{\mathcal{C}}_T&=\frac{\Omega}{2d(d+2)}[(d-1)(d+2){\mathcal{A}}-2{\mathcal{B}}-4(d+1){\mathcal{C}}]\,, \\
 t_2&=\frac{2(d+1)}{d}\frac{(d-2)(d+2)(d+1){\mathcal{A}}+3d^2{\mathcal{B}}-4d(2d+1){\mathcal{C}}}{(d-1)(d+2){\mathcal{A}}-2{\mathcal{B}}-4(d+1){\mathcal{C}}}\,,\\
t_4&=-\frac{d+1}{d}\frac{(d+2)(2d^2-3d-3){\mathcal{A}}+2d^2(d+2){\mathcal{B}}-4d(d+1)(d+2){\mathcal{C}}}{(d-1)(d+2){\mathcal{A}}-2{\mathcal{B}}-4(d+1){\mathcal{C}}}\,.
\end{split}
\end{align}
}
The structure of higher point functions of the stress tensor are less understood, see however~\cite{Dymarsky:2013wla}. There is an intriguing connection between the allowed structures in $n$-point functions, and the independent vertices in a $d+1$ dimensional gravitational problem~\cite{Hofman:2008ar,Costa:2011mg}.

I expect that by writing down all the higher derivative terms not containing derivatives of the curvature, I account for a significant fraction of the possible gravity vertices in AdS. Then by AdS/CFT, or by the counting argument of~\cite{Costa:2011mg}, I expect to capture a large range of possible behaviors for CFT $n$-point functions. I only work  to linear order in the higher curvature couplings, and use higher derivative gravity in the above kinematical sense. Hence, I do not assume any sophisticated consistency condition about the resulting theory.

\pagebreak

From the perturbative approach to EE in field theory, I expected that the answer in the holographic computation would depend on the three parameters determining the three-point function. My approach is sensitive to all higher point functions as well, hence I would be able to detect, if the result depended on coefficients that only appear in higher point functions. Instead, I find that only $C_T$ appears in the final result~\eqref{EinsteinArea}. My result then points towards a natural first step for field theory attempts to evaluate the quantity $s_d^{(2)}(\Sig)$~\eqref{sExpansion}: one should show that the tensor structures multiplying $t_2,\, t_4$ do not contribute. Alternatively, if the perturbative approach in its current incarnation only works for R\'enyi entropies, not EE (as suggested by the failure in~\cite{Rosenhaus:2014zza}), my result provides clues for the correct analytic continuation.

I would like to add that field theory computations -- similar in philosophy to this current one -- were done for ${d\ov dq} S_q\big\vert_{q=1}$~\cite{Perlmutter:2013gua} and  ${d^2\ov dq^2} S_q\big\vert_{q=1}$~\cite{Hung:2014npa,Lee:2014zaa}. These quantities are also controlled by two- and three-point functions of the stress tensor. Consequently, ${d\ov dq} S_q\big\vert_{q=1}$ is determined by $C_T$, and ${d^2\ov dq^2} S_q\big\vert_{q=1}$ depends on all three of $C_T,\, t_2,\, t_4$. My computation is the analog of this latter case, but my result does not depend on $t_2,\, t_4$. 

Intriguingly,~\cite{Nozaki:2013vta} obtained an integral expression for $s_d^{(2)}(\Sig)$ in Einstein gravity, which is equivalent to~\eqref{EinsteinArea}. They speculate that their formula is an integrated stress tensor two-point function. It would be very interesting to see the field theory derivation of $s_d^{(2)}(\Sig)$.

\section{EE of a deformed sphere in higher curvature gravity}\label{sec:Gravity}

\subsection{EE in higher derivative gravity}\label{sec:Dong}

Using the generalized gravitational entropy introduced in~\cite{Lewkowycz:2013nqa},~\cite{Dong:2013qoa} determined a nice prescription for the EE in higher derivative gravity, whose Lagrangian does not contain derivatives of the Riemann tensor.\footnote{\cite{Camps:2013zua} obtained less general results with similar techniques at the same time. There has been recent progress on theories with derivatives of the Riemann tensor~\cite{Miao:2014nxa}.} I review the procedure below. EE is given by evaluating the generalized area functional~\eqref{eei} on the (generalized) Ryu--Takayanagi (RT) surface~\cite{Ryu:2006bv}.  In Einstein gravity the RT surface is the minimal area surface homologous to the boundary region $V$. In a higher curvature theory  it is not known, if the equation of motion for the RT surface can be obtained from the minimization of the generalized entropy functional~\eqref{eei}. See~\cite{Dong:2013qoa,Bhattacharyya:2014yga} for discussion of this point. This issue does not affect my work, as I discuss in Sec.~\ref{sec:Surface}.

 Firstly, one has to go to coordinates adapted to the surface:
\es{met3}{
ds^2 =& \[dz d\zb + T (\zb dz-z d\zb)^2 \] + \(g_{ij} + 2K_{aij} x^a + Q_{abij} x^a x^b\) dy^i dy^j \\
&+ 2i \(U_i + V_{ai} x^a\) \(\zb dz-z d\zb\) dy^i + \cdots \,.
}
These coordinates can be constructed by shooting geodesics normal to the surface, thereby getting the transverse coordinates  to the surface, $x^a=(z,\, \zb)$, and constructing a set of coordinates along the surface, $y^i$. 

EE is then given by the following functional evaluated on the RT surface:
\be\la{eei}
S_{EE}= 2\pi\int d^{d-1} y \sqrt{g} \lt\{ \fr{\pa \sL}{\pa R_{z\zb z\zb}} + \sum_\a \(\fr{\pa^2 \sL}{\pa R_{zizj} \pa R_{\zb k\zb l}}\)_\a \fr{8K_{zij} K_{\zb kl}}{q_\a+1} \rt\} \,.
\ee
I will refer to the first term of the sum as Wald, to the second as Dong term. The second derivative of $\sL$ has to be written in components.  The following components of the Riemann tensor have to be further expanded in terms of the extrinsic curvature $K_{aij}$, $Q_{abij} \eq \pa_a K_{bij}$, and the lower-dimensional Riemann tensor $r_{ikjl}$:
\ba
R_{abij} &= \td R_{abij} + g^{kl} (K_{ajk} K_{bil} - K_{aik} K_{bjl}) \,,\nn\\
R_{aibj} &= \td R_{aibj} + g^{kl} K_{ajk} K_{bil} - Q_{abij} \,,\la{rexpi}\\
R_{ikjl} &= r_{ikjl} + g^{ab} (K_{ail} K_{bjk} - K_{aij} K_{bkl}) \,,\nn
\ea
where I have defined
\es{rtd}{
\td R_{abij} &\eq 2 \ve_{ab} (\pa_i U_j - \pa_j U_i) \,,\\
\td R_{aibj} &\eq \ve_{ab} (\pa_i U_j - \pa_j U_i) + 4 g_{ab} U_i U_j  \,,
}
where $\ve_{z\zb}=-\ve_{\zb z}=i/2$. I will use $\a$ to label the terms in the expansion.  For each term in the expansion of the second derivative of $\sL$, which is itself a product, $q_\a$ is defined as the total number of $Q_{zzij}$ and $Q_{\zb\zb ij}$, plus one half times the total number of $K_{aij}$, $R_{abci}$, and $R_{aijk}$.  Finally, one has to sum over $\a$ with weights $1/(1+q_\a)$.  

As an example I consider the Lagrangian:
\es{4DerivLag}{
S =& -\fr{1}{16\pi G_N} \int d^{d+1}x \sqrt G \le[R+{6\ov L^2} +\l_1 L^2 R^2 + \l_2 L^2 R_{\m\n} R^{\m\n} + \l_3 L^2 R_{\m\r\n\s} R^{\m\r\n\s}\ri]\,,
}
where $G_{\mu\nu}$ denotes the full $d+1$ dimensional metric.~\cite{Fursaev:2013fta,Dong:2013qoa} determined a closed form expression for the EE of this theory from~\eqref{eei}:
\es{4DerivativeGeneral}{
S_{EE}= {1\ov 4 G_N}\int d^{d-1} y \sqrt{g} \[1+ 2\l_1 L^2 R + \l_2 L^2 \(\T Raa -\fr12 K_a K^a\) + 2\l_3 L^2 \(\T R {ab}{ab} - K_{aij} K^{aij}\) \]
}
I will get away without determining such an expression for the case of a general $\sL$ and still compute EE to $\sO(\ep^2)$ by using the special features of my problem. 

\subsection{Stress tensor correlators in higher derivative gravity}

Before embarking on the calculation of the deformed sphere EE, I review the relevant results of the very useful paper~\cite{Sen:2014nfa}, which are based on the background field formalism~\cite{Miao:2013nfa}. The prespriction is to treat $G_{\m\n}$ and $R_{\m\n\r\s}$ as independent variables, and expand $ {\mathcal{L}}(G^{\m\n},R_{\m\n\r\s})$ in $R_{\m\n\r\s}$ around the quantity $\bar{R}_{\m\n\r\s}=-{1\ov \tL^2}\le(G_{\m\r} G_{\n\s}-G_{\m\s} G_{\n\r}\ri)$, where $\tilde{L}$ is the AdS radius. I denote the difference of the curvature from this background tensor by
\es{BackgroundDiff}{
\D R_{\m\n\r\s}=R_{\m\n\r\s}-\bar{R}_{\m\n\r\s}\,.
}
Note that $\D R_{\m\n\r\s}=0$ on AdS, as $R_{\m\n\r\s}=\bar{R}_{\m\n\r\s}$ when $G_{\m\n}=G^\text{(AdS)}_{\m\n}$.

The Lagrangian in the background field expansion takes the form
\be\label{lag}
{\mathcal{L}}=-{1\ov 16\pi G_N}\le({c_0\ov \tL^2}+c_1\D R+\frac{c_4 \tL^2}{2} \D R^2+\frac{c_5\tL^2}{2} \D R_{\m\n} \D R^{\m\n}+\frac{c_6\tL^2}{2} \D R_{\m\n\r\s}\D R^{\m\n\r\s}+\sum_{i=1}^{8}\tilde{c}_i \tL^4\D {\mathcal{K}}_i+\cdots\ri),
\ee
where $c_0=-2d c_1$ and I used the notation $\D {\mathcal{K}}_{i}={\mathcal{K}}_{i}|_{R\rightarrow \D R}$. The basis for the third order terms is given by 
\es{Ki}{
{\mathcal{K}}_i=&(R^3,\, R^\m_{\ \n} R^\n_{\ \r} R^\r_{\ \m},\, R R^{\m\n}R_{\m\n},\, R R^{\m\n\r\s}R_{\m\n\r\s},\, R^{\m\n}R^{\r\s}R_{\m\s\r\n\s},\, \\
&R_{\m\n}R^{\m \r\s\l}R^{\n}_{ \ \r\s\l},\, R_{\m\n\r\s}R^{\m\n \l\d}R^{\r\s}_{ \ \ \l\d},\, R_{\m\n\r\s}R^{\m \l \r\d}R^{\n \  \s}_{ \ \l \ \d} )\,.
}

For the four derivative theory~\eqref{4DerivLag}, the coefficients in~\eqref{lag} are given by:
\es{ExplicitCoeff}{
c_1&=1-2 \ga\le(d(d+1)\l_1+d\l_2+2\l_3\ri)\,,\\
c_4&=2\ga\l_1\,, \qquad c_5=2\ga\l_2 \,, \qquad c_6=2\ga\l_3\,, \\
\tilde c_i&=0\,,
}
where I introduce $\ga\equiv L^2/\tL^2$ that determines the relationship between the cosmological constant and the AdS radius. $\ga$ is determined by the equation:
\es{gaEq}{
0=1-3\ga+\ga^2\, {d-3\ov d-1} \le(d(d+1)\l_1+d\l_2+2\l_3\ri)\,.
}

\cite{Sen:2014nfa} calculates $C_T,\, t_2,\, t_4$ in this formalism. In terms of the coefficients in~\eqref{lag} they are given by:
\es{CTBackground}{
C_T&= {\tilde{L}^{d-1}\ov  4 G_N}\, \frac{\Gamma(d+2)}{2\pi^{d+2\ov 2}(d-1)\Gamma\le(d\ov 2\ri)}\, \le(c_1+2(d-2)c_6\ri)\,,\\
t_2 &={\tilde{L}^{d-1}\ov 16\pi G_N}\,\frac{d(d - 1)}{c_1+2(d-2)c_6} [2c_6-12 (3d + 4) \tilde c_7+3 (7 d + 4) \tilde c_8]\,,\\
t_4&= {\tilde{L}^{d-1}\ov 16\pi G_N}\,\frac{6d(d^2 - 1)(d+2)}{c_1+2(d-2)c_6}(2\tilde c_7 -\tilde c_8)\,.
}

A nice consistency check of the above formulae is that in $d+1=4$ bulk dimensions I get no contribution from the Gauss--Bonnet term, $R^2-4 R_{\m\n} R^{\m\n} +  R_{\m\r\n\s} R^{\m\r\n\s} $. The reason is that in four dimensions the Gauss--Bonnet term is a topological, hence it should not contribute to the propagation of gravitons that determine $C_T, \, t_4$. Note that $t_2$ does not have a meaning in $d=3$, correspondingly the stress tensor three-point function only depends on two parameters in a parity invariant theory.

\subsection{Ryu--Takayanagi surface in Einstein gravity}\label{sec:Surface}

I will only need some pieces of the general framework presented in Sec.~\ref{sec:Dong}. I will only work to first order in the higher derivative couplings. This means that I can use the RT surface of Einstein gravity, and evaluate~\eqref{eei} on this surface. The surface is corrected at first  order in the higher derivative couplings, but as usual they do not change the on-shell action $S_{EE}$ to linear order. 

Let me specify the particular features of the problem. Firstly, I describe the RT surface in a coordinate system that {\it is not} the coordinate system in~\eqref{met3}. The usual parametrization of AdS$_{d+1}$ in Poincar\'e coordinates is
\es{PoincareAdS}{
ds^2= {\tL^2\ov u^2}\le(dt^2+ dr^2+r^2d\Om_{d-2}^2+du^2\ri) \,.
}
I use $\tL$ to denote the true AdS radius, which does not necessarily agree with $L$ determining the cosmological constant as in~\eqref{4DerivLag}, due to corrections coming from higher derivative terms. I use $u$ instead of the usual $z$ to avoid the clash of notation with the surface adapted coordinates~\eqref{met3}. I then trade $(r,\, u)$ for $(\rho,\, \Theta)$ by
\es{CoordTf}{
\rho&=\sqrt{r^2+u^2}\,,\\
\Theta&=\arctan {r\ov u}\,.
}
In these coordinates
\es{PoincareAdS2}{
ds^2= {\tL^2\ov \rho^2\cos^2\Theta}\le(dt^2+ d\rho^2+\rho^2d\Omega_{d-2}^2\ri) \,,
}
where $\Omega_{d-1}=(\Theta, \Omega_{d-2})$ and $\Omega_{d-2}$ represents the field theory angular directions. Note that the range of $\Theta$ is only $[0,\pi/2]$, so the constant $\rho$ slices are hemispheres. The AdS boundary is at $\Theta=\pi/2$.

The surface computing the EE of a circle is a hemisphere mentioned above. The RT surface for a deformed sphere is given by:
\es{Surface}{
\rho(\Theta,\Omega_{d-2})&=1+\ep \, \rho_1(\Theta,\Om_{d-2})+\sO(\ep^2)\,.
}
Naively, the  ${\cal O}(\ep^2)$ correction would also have to be kept, but I found that it does not contribute to the quantities appearing in~\eqref{met3}. It would be good to understand this cancellation better. 
By plugging in~\eqref{Surface} into the equation of motion coming from the area functional (relevant for Einstein gravity), and expanding in $\ep$, I obtained the surface anchored on~\eqref{GeomRegion} in general dimensions:
\es{Surface2}{
\rho_1(\Theta,\Omega_{d-2})&=\sum_{\ell, m_1,\dots, m_{d-3}} a_{\ell, m_1,\dots, m_{d-3}} Y_{\ell, m_1,\dots, m_{d-3}}(\Omega_{d-2}) f_\ell(\Theta)\,,\\  
f_\ell(\Theta)&\equiv C_\ell \, \sin ^{\ell}(\Theta ) \,\, _2F_1\left(\frac{\ell-1}{2},\frac{\ell}{2};\frac{1}{2} (d+2 \ell-1);\sin ^2(\Theta )\right)\,,\\
C_\ell &\equiv \frac{\sin \left(\frac{\pi  d}{2}\right) \Gamma \left(1-\frac{d}{2}\right) \Gamma \left(\frac{1}{2} (d+\ell-1)\right) \Gamma \left(\frac{d+\ell}{2}\right)}{\pi  \Gamma
   \left(\frac{1}{2} (d+2 \ell-1)\right)} \,.
}
This minimal surface was also derived in~\cite{Nozaki:2013vta}, where an equivalent integral expression is given.
Note that different spherical harmonics do not mix, and to this order in $\ep$ the solution is just a superposition.~\eqref{Surface2} simplifies in odd $d$: in $d=3$ the answer for a deformed circle is given by~\cite{Hubeny:2012ry,Allais:2014ata}:
\es{Surface3}{
\rho_1(\Theta,\Phi)&=\sum_{\ell}\le(a^{(c)}_\ell\, {\cos\le(\ell\Phi\ri)\ov \sqrt\pi}+a^{(s)}_\ell\, {\sin\le(\ell\Phi\ri)\ov \sqrt\pi}\ri)\, \tan^\ell {\Theta\ov2} \le(1 + \ell \cos \Theta\ri) \,.
}
where $\Phi$ is the field theory angular direction; in $d=5$, $f_\ell(\Theta)$ takes the form:
\es{Surface4}{
f_\ell(\Theta)=\tan ^{\ell}\, \frac{1+(\ell+1) \cos \Theta +{\ell (\ell+2) \ov 3}\cos ^2\Theta  }{1+\cos \Theta}\,.
}

\subsection{Simplification of the EE functional}

The next step is to determine the coordinates adapted to the surface~\eqref{Surface} as in~\eqref{met3}. I have to construct geodesics emanating from the surface. I denote the affine parameter of the geodesics by $s$ ($s=0$ at the surface), the angle of their tangent vector to the vector $\p_t$ by $\sin \tau$, and the starting point on the surface by $\omega_{d-1}=(\theta, \omega_{d-2})$. I construct the geodesics by solving the geodesic equation in a power series in $s$, thereby covering a small neighborhood of the surface with coordinates $(s,\tau, \omega_{d-1})$.  For illustration purposes, I show the linear in $s$ piece to ${\cal O}(\ep)$:
\es{Geodesic}{
\begin{pmatrix}
t\\
\rho\\ 
\Omega_{d-1} 
 \end{pmatrix}
 =
 \begin{pmatrix}
s  \sin\tau \cos\theta \le[1 + \ep \rho_1( \omega_{d-1})\ri]\\
 1 + \ep \rho_1(\theta, \phi) +   s \cos\tau \cos\theta  \le[1 + \ep \rho_1(\omega_{d-1})\ri]\\ 
 \omega_{d-1}- s \,\ep \cos\tau \cos\theta \, \p_{\omega_{D-2}}\rho_1( \omega_{d-1})
 \end{pmatrix}
 +{\cal O}(\ep^2,s^2) \,,
 }
where $\p_{\omega_{d-1}}$ is a shorthand for $\tilde{g}^{ij}\p_i$, where $\tilde{g}_{ij}$ is the round $S^{d-1}$ metric. I need these expressions up to ${\cal O}(\ep^2,s^3)$ to do the calculation, but the expressions are too lengthy to print.
 Finally, I change variables to
 \es{zzb}{
 z&=s\,e^{i\tau}\,,\\
 \zb&=s\,e^{-i\tau}\,,
 }
to get to the final adapted coordinate system with $x^a=(z,\zb)$ and $y^i=\omega_{d-1}$. I can now go ahead and read off the functions that appear in~\eqref{met3}. It is worth displaying $g_{ij}$ here:
\es{gDisp}{
ds^2\big\vert_{z=\zb=0}&={\tL^2\ov \cos^2\theta}\le[\tilde{g}_{ij}+\ep^2\p_i\rho_1(y)\,\p_j\rho_1(y)+\sO(\ep^3)\ri] dy^i dy^j \,.
}
 Note that there is no linear in $\ep$ piece in $g_{ij}$. To construct all the curvatures needed to evaluate~\eqref{eei} I use that AdS is a maximally symmetric space, hence:
\es{MaxSym}{
R_{\m\n\r\s}=-{1\ov \tL^2}\le(G_{\m\r} G_{\n\s}-G_{\m\s} G_{\n\r}\ri) \,.
}
I will also need $K_{aij}$. To leading order in $\ep$ it can be constructed from starting with the adapted  coordinates of the sphere:
\es{KDisp}{
K_{zij}&=K_{\zb ij}=\ha\le[\nabla_i n_j-n_i n^\mu \nabla^{(0)}_\mu n_j\ri]+\sO(\ep^2) \,, \\
n_\mu&=\le(\ha,\ha,\, -\delta_{i}^\theta \, {z+\zb\ov 2}\tan\theta-{\ep\ov \cos\theta} \p_i \rho_1(y) \ri)\,,
}
where the $(0)$ superscript denotes covariant derivatives compatible with the metric at $\ep=0$, i.e.~the metric adapted to the sphere.

There are several remarks in order:
\ben[(i)]
\item Because the surface only differs for a hemisphere at ${\cal O}(\ep)$, the extrinsic curvature $K_{aij}={\cal O}(\ep)$. This brings about a major simplification; because I only want the answer to  ${\cal O}(\ep^2)$, I only need to evaluate the second derivative of $\sL$ in the Dong term of~\eqref{eei} to zeroth order in $\ep$. 

\item $U_i$ can be thought of as a Kaluza--Klein gauge field. In AdS it is a flat connection, and my choice of coordinates makes it zero. Hence, $ \td R_{abij}, \td R_{aibj}$ vanish.

\item Making use of the above two points, I can make the replacements in~\eqref{rexpi}:
\ba
R_{abij} &\to 0 \,,\nn\\
R_{zi\zb j} &\to - Q_{z\zb ij} \,, \quad R_{ziz j} \to 0\,, \quad R_{\zb i \zb j} \to 0 \,,\\
R_{ikjl} &\to r_{ikjl}  \,,\nn
\ea
where the RHS is understood to be true at $\sO(\ep^0)$. Hence, these replacements should only be used in the Dong term. For the rest of the Riemann tensor components I can use~\eqref{MaxSym} evaluated on the surface. Because the metric is block diagonal on the surface $R_{abci}$ and $R_{aijk}$ are all zero.\footnote{Due to the vanishing of $U_i$ (and $V_{ai}$), the metric is block diagonal even away from the surface to the order exhibited in~\eqref{met3}.}  Thus, I remarkably get that all terms that would contribute to the count in $q_\al$ are zero. Hence, $q_\a=0$ for all the nonzero terms in the sum. 

\een

In the four-derivative case, I already have the formula~\eqref{4DerivativeGeneral}, that is easily evaluated. For the most general $\sL$, however, the above remarks bring major simplification.
From what I explained above, I conclude that~\eqref{eei} can be replaced by:
\es{eei2}{
S_{EE}&=S_\text{Wald}+S_\text{Dong}\,,\\
 S_\text{Wald}&=2\pi\int d^{d-1} y\ \sqrt{g} \ \fr{\pa \sL}{\pa R_{z\zb z\zb}}\,, \\
S_\text{Dong}&=16 \pi\int d^{d-1} y\ \le[\sqrt{g} \ \fr{\pa^2 \sL}{\pa R_{zizj} \pa R_{\zb k\zb l}}\ri]_{\ep=0} \, K_{zij} K_{\zb kl} \,.
}
Firstly, I will analyze the Wald term. 
From the background field Lagrangian~\eqref{lag} it is easy to see that 
\es{BackWald}{
16\pi G_N \,\sqrt{g}\ \fr{\pa \sL}{\pa R_{z\zb z\zb}} =-{c_1 \sqrt{g} \ov 2}\le(G^{zz}G^{\zb\zb}-G^{z\zb}G^{z\zb}\ri)=2c_1\sqrt{g}\,.
}
Secondly, I look at the Dong term. Again, from the background field Lagrangian it is clear that only the terms quadratic in $\D R_{\m\n\r\s}$ contribute. A short calculation gives:
\es{BackDong}{
16\pi G_N\,\le[\sqrt{g} \ \fr{\pa^2 \sL}{\pa R_{zizj} \pa R_{\zb k\zb l}}\ri]_{\ep=0} \, K_{zij} K_{\zb kl} =-\sqrt{g}\big\vert_{\ep=0}\, \le[{c_5 \tL^2\ov16}\,K_{a}K^{a}+ {c_6\tL^2\ov 4}\, K_{aij}K^{aij}\ri]\,.
}
A nice sanity check of~\eqref{BackWald} and~\eqref{BackDong} is to compare them to~\eqref{4DerivativeGeneral} with $c_i$ given in~\eqref{ExplicitCoeff}. The curvature terms combine exactly into $c_1$ and the coefficients of the extrinsic curvature terms also match.\footnote{There is also a more elementary argument for the structure of this result that does not use the background field technique. Note that the Wald term is an arbitrary combination of Riemann tensors that has four uncontracted indices. Using~\eqref{MaxSym} these fall apart into metric tensors. Because the metric is block diagonal, the only pairing of indices is $(G^{z\zb})^2=4$. A similar argument applies to the Dong term. However, this method does not enable us to keep track of the overall coefficients effectively. 

From these elementary considerations it follows that the Wald term vanishes, if there are at least two Weyl tensors in $\sL$, while the Dong term vanishes, if there are at least three. The argument is simple, the Weyl tensor vanishes in AdS, as it is a conformally flat space. Hence, unless the derivatives hit all the Weyl terms, the result will be zero. If the field theory argument that $s_d^{(2)}$ depends on the integrated stress tensor three-point function gets elevated to a proof in the future, then one can think of this work as a trick that uses holography to evaluate the integrated three-point function. In $d=3$ from the above argument I immediately conclude that the $s_3^{(2)}$ cannot depend on $t_4$, as adding a Weyl cubed term to $\sL$ changes $t_4$, while $s_3^{(2)}$ remains unchanged. The full form of the shape dependence can then be fixed from one CFT example, i.e.~from the dual of Einstein gravity. With specific choices of $\sL$, in higher dimensions the $t_2,\, t_4$ independence can also be established. However, in the absence of better field theory understanding I proceed without making the assumption that the result can only depend on $C_T,\, t_2,\, t_4$.}

\pagebreak

The minimal surface condition in Einstein gravity results in the equation of motion:
\es{EomK}{
K_a=0 \,.
}
Because the surface~\eqref{Surface} satisfies this equation at every order in $\ep$, I can further simplify the Dong term. Finally, I obtain
\es{DongResult}{
S_{EE}&={1\ov 4 G_N}\int d^{d-1} y \ \le[c_1\sqrt{g}-c_6\tL^2\sqrt{g}\big\vert_{\ep=0}\, K_{aij}K^{aij} \ri]\,.
}

\subsection{Evaluation of the EE of a deformed sphere}

Finally, I have to evaluate~\eqref{DongResult}. The first term, $\int d^{d-1} y \ \sqrt{g}$ is just the area of the RT surface that would give the entropy in Einstein gravity. I could have evaluated this integral in Sec.~\ref{sec:Surface}. However, it is more convenient to evaluate it in the surface adapted coordinates~\eqref{gDisp}, as from the form of $g_{ij}$~\eqref{gDisp} I see right away that the $\sO(\ep^2)$ correction to the shape of the surface does not contribute to the area. This is rather hard to see in other coordinate systems. I expand $\sqrt{g}$ as:
\es{VarSqrt}{
\sqrt{g}={\sqrt{\tilde{g}}\ov \cos^{d-1}\theta}\le[1+{\ep^2\ov 2}{\cos^2\theta}\,  \le(\tilde{g}^{ij}\p_i \rho_1\p_j \rho_1\ri)\ri] \,,
}
where $\tilde{g}_{ij}$ is the round metric on $S^{d-1}$, as before.
Evaluating the area through explicit calculation gives the contribution to the universal term:
\es{EinsteinArea2}{
&\le[\int d^{d-1} y \ {\sqrt{\tilde{g}}\ov 2 \cos^{d-3}\theta}  \le(\tilde{g}^{ij}\p_i \rho_1\p_j \rho_1\ri)\ri]_\text{univ}=\\
&\qquad\qquad\qquad\qquad =\tL^{d-1}\sum_{\ell, m_1,\dots, m_{d-3}}  {a_{\ell, m_1,\dots, m_{d-3}}^2\ov 2^{d-1} \Gamma\le(d\ov2\ri)^2}\prod_{k=1,\dots d} (\ell+k-2)\times \begin{cases}
{\pi\ov 2} \qquad &\text{$d$ odd,}\\
1 \qquad &\text{$d$ even.}
\end{cases}
}
\cite{Nozaki:2013vta} obtained an equivalent integral expression. 

The evaluation of the simplified Dong term again proceeds through explicit calculation. Although the integrand is rather dissimilar to the Wald term, the universal contribution has the same shape dependence.\footnote{The divergent contributions differ. This demonstrates that the two terms cannot be deformed into each other using any clever manipulations.} It would be good to understand why this is the case from a general argument. By adding the Wald and Dong terms, I get the following combination multiplying~\eqref{EinsteinArea2}:
\es{Combiantion}{
{c_1+2(d-2)c_6\ov 4 G_N}= \frac{2\,\pi^{d+2\ov2}(d-1)\Gamma\le(d\ov2\ri)}{\Gamma(d+2)}{C_T\ov \tL^{d-1}}\,,
}
where I used~\eqref{CTBackground}. Note that I was only working to linear order in the higher curvature couplings, so the agreement valid to all orders in the couplings is a nice coincidence.

Thus, the final result is:
\es{EinsteinArea3}{
s_d^{(2)}(\Sig)&=C_T\,  {\pi^{d+2\ov2}(d-1)\ov 2^{d-2} \Gamma(d+2)\Gamma\le(d\ov2\ri)}\sum_{\ell, m_1,\dots, m_{d-3}} a_{\ell, m_1,\dots, m_{d-3}}^2 \prod_{k=1,\dots d} (\ell+k-2)\times \begin{cases}
{\pi\ov 2} \qquad &\text{$d$ odd,}\\
1 \qquad &\text{$d$ even,}
\end{cases}
}
as announced in the introduction.

\section*{Acknowledgments}
I thank I.~Klebanov, A.~Lewkowycz, R.~Myers, E.~Perlmutter, S.~Pufu, A.~Sinha, M.~Smolkin, C.~von Keyserlingk, and especially X.~Dong for useful discussions and comments on the draft. During the exploratory stages of this project I greatly benefitted of the Mathematica package~\cite{xAct}. I am supported by the Princeton Center for Theoretical Science.

\bibliographystyle{ssg}
\bibliography{shape}

\end{document}